\documentstyle[preprint,aps,epsf,floats]{revtex}
\begin{document}

\draft
 
{\tighten
\preprint{\vbox{\hbox{CALT-68-2097} 
                \hbox{MIT-CTP-2611} 
                \hbox{hep-ph/9702322} }}

\title{Comment on nonperturbative effects in $\bar B\to X_s\,\gamma$}
 
\author{Zoltan Ligeti$\,^a$, Lisa Randall$\,^b$, and Mark B.\ Wise$\,^a$}
 
\address{ \vbox{\vskip 0.truecm}
  $^a$California Institute of Technology, Pasadena, CA 91125 \\[6pt]
  $^b$Massachusetts Institute of Technology, Cambridge, MA 02139  }

\maketitle 
\widetext 

\begin{abstract}%
Uncertainties in the theoretical prediction for the inclusive 
$\bar B\to X_s\,\gamma$ decay rate are discussed.  We emphasize that there is
no operator product expansion for this process.  Nonetheless, some
nonperturbative effects involving a virtual $c\,\bar c$ loop are calculable
using the operator product expansion.  They give a contribution to the decay
rate that involves the $B$ meson matrix element of an infinite tower of
operators.  The higher dimension operators give effects that are only
suppressed by powers of $m_b\Lambda_{\rm QCD}/m_c^2\sim0.6$, but come with
small coefficients.

\end{abstract}

}

\newpage

The inclusive $\bar B\to X_s\,\gamma$ decay has received considerable attention
in recent years [1--15],
since it is sensitive to physics beyond the standard model \cite{np} ($X_s$
denotes a final state with strangeness $-1$ and charm 0).  The photon spectrum
also carries information on nonperturbative physics that can help us better
understand other $B$ decays \cite{Matthias,AKZL}.  The recent CLEO measurement
\cite{CLEO} excludes large deviations from the standard model.  Therefore, it
is important to know the standard model predictions as precisely as possible.

Since the $b$ quark is heavy compared to the QCD scale, one would hope that the
inclusive $\bar B\to X_s\,\gamma$ decay rate can be calculated in a systematic
QCD-based expansion \cite{CGG}.  The dominant contribution to the decay rate
comes from the matrix element of the electromagnetic penguin operator (usually
denoted by $O_7$).  In the $m_b\to\infty$ limit, it is given by the free quark
decay result.  The leading nonperturbative corrections to this contribution are
suppressed by $(\Lambda_{\rm QCD}/\,m_b)^2$.  Provided the photon energy is not
restricted to be too close to its maximal (i.e., end-point) value, they are
quite small, around $-3$\% \cite{FLS}.  With the recent completion of the full
next-to-leading order perturbative calculation \cite{nlo}, it is usually argued
that theoretical uncertainties in the prediction for the inclusive $\bar B\to
X_s\,\gamma$ decay rate are not larger than 10\%.  

The effective weak interaction Hamiltonian at a scale $\mu$ (of order $m_b$)
is given by
\begin{equation}
H_{\rm eff} = -{4G_F\over\sqrt2}\, V_{ts}^*\,V_{tb}\, 
  \sum_{i=1}^8 C_i(\mu)\, O_i(\mu) \,.
\end{equation}
In the conventional notation, 
$O_2=(\bar s_{L\alpha}\,\gamma_\mu\,b_{L\beta})\,
(\bar c_{L\beta}\,\gamma^\mu\,c_{L\alpha})$,
$O_1$ only differs from $O_2$ in the way color indices are contracted,
$O_3-O_6$ are four-quark operators involving all flavors below the scale $\mu$,
$O_7=(e/16\pi^2)\,m_b\,\bar s_L\,\sigma^{\mu\nu} F_{\mu\nu}\,b_R$, and 
$O_8$ is obtained from $O_7$ by replacing $eF_{\mu\nu}$ by $g_sG_{\mu\nu}$.
Using perturbative QCD to evaluate the $\bar B\to X_s\,\gamma$ decay rate,
in the leading logarithmic approximation the matrix element of 
$C_7(\mu)\,O_7(\mu)$ dominates for large enough photon energies.  

A systematic computation of the $O_7$ contribution to the inclusive $\bar B\to
X_s\,\gamma$ decay rate involves performing an operator product expansion (OPE)
for the time ordered product 
\begin{equation}\label{ope1}
T_{77} = {i\over2m_B}\,\int {\rm d}^4x\, e^{-iq\cdot x}\, \langle\bar B(v)|\, 
  T\{O_7^{\mu\dagger}(x)\, O_7^\nu(0)\} |\bar B(v)\rangle\, g_{\mu\nu} \,,
\end{equation}
to all orders in the strong interaction.  Here
$O_7^\mu=(i\,e/8\pi^2)\,m_b\,\bar s_L\,\sigma^{\mu\lambda}q_\lambda\,b_R$.  At
fixed $q^2=0$, this time ordered product has a cut in the complex $v\cdot q$
plane along $v\cdot q<m_b/2$ corresponding to final hadronic states $X_s$, and
another cut along $v\cdot q>3m_b/2$ corresponding to final hadronic states
$X_{bb\bar s}$.  The contribution of the magnetic moment operator $O_7$ to the
$\bar B\to X_s\,\gamma$ decay rate is given by the discontinuity across the cut
in the region $0<v\cdot q<m_b/2$,
\begin{equation}\label{77spec}
{{\rm d}\Gamma\over{\rm d}E_\gamma} = {4G_F^2\, |V_{ts}^*\,V_{tb}|^2\, C_7^2
  \over \pi^2}\, E_\gamma\, {\rm Im}\, T_{77}\,.
\end{equation}
Since the cuts are well-separated, one can compute this contribution to the
$\bar B\to X_s\,\gamma$ decay rate assuming local duality at the scale $m_b$. 
(The integration of $T_{77}$ over the contour $C$ in Fig.~1 pinches the
physical cut at $v\cdot q=0$, but at that point the hadronic final states have
invariant mass $m_{X_s}=m_B$.) 

\begin{figure}[t] 
\centerline{\epsfysize=8truecm \epsfbox{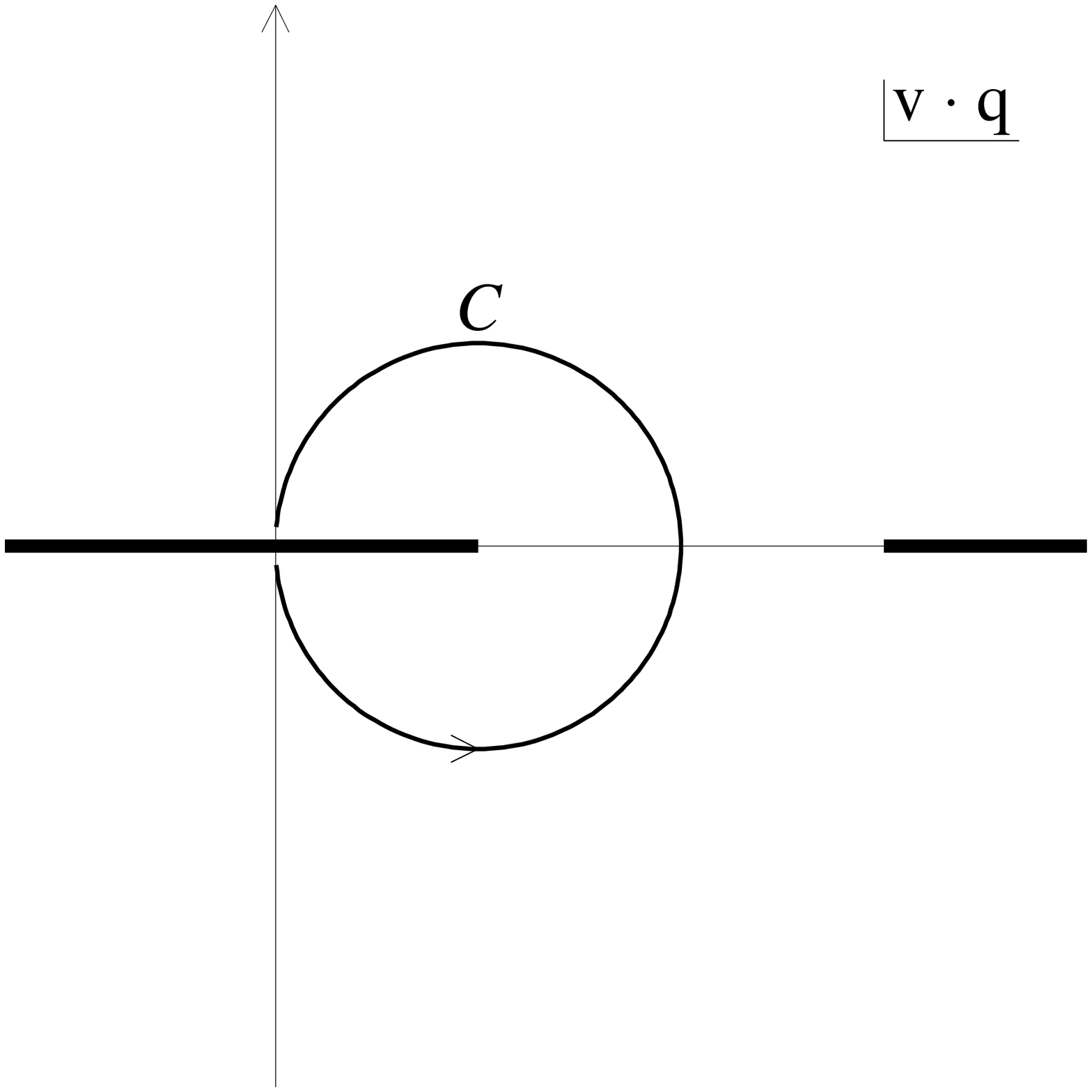}}
\caption[1]{The integration contour $C$ in the complex $v\cdot q$ plane.
The cuts extend to ${\rm Re}\,v\cdot q\to\pm\infty$. }
\end{figure}

At leading order in the OPE, the dimension-three operator $\bar
b\,\gamma_\mu\,b$ occurs.  Its matrix element gives a calculable contribution
to the photon energy spectrum proportional to $\delta(E_\gamma-m_b/2)$.  Higher
dimension operators give terms proportional to derivatives of this delta
function, and the matrix elements of the operators with dimension greater than
five are not known.  In order to justify retaining only the lowest dimension
operators, the photon energy must be averaged over a region $\Delta
E_\gamma\gg\Lambda_{\rm QCD}$.  At the present time these higher dimension
operators introduce a significant uncertainty, since the photon spectrum is
only measured experimentally over a region about $500\,$MeV from the end-point
\cite{CLEO}.

When operators in $H_{\rm eff}$ other than $O_7$ are included, the $\bar B\to
X_s\,\gamma$ decay receives contributions from diagrams in which the photon
couples to light quarks.  It is well-known that for such processes, there are
uncalculable contributions suppressed by $\alpha_s$, but not by powers of the
scale associated with the process.  Typically, the leading logarithms are
calculable \cite{ed}, but terms suppressed by a logarithm (or equivalently by
$\alpha_s$, but not by a power) can only be estimated using information on the
fragmentation functions $D_{q\to\gamma X}$ and $D_{g\to\gamma X}$ deduced from
other experiments or from models.  While this may be worrisome, experience
shows that usually the leading order perturbative QCD calculation provides an
order of magnitude estimate of these effects \cite{lore}.  Perturbative
computations indicate that for weak radiative $B$ decays into hard photons both
the contribution of light quark loops \cite{lql}, and the effects related to
decay functions of light partons into a photon \cite{KLP}, are very
small.\footnote{For soft photons these effects are important.  There are also
interference effects where the photon couples to a light quark and to the charm
quark, or to a light quark and through $O_7$.  These are also small for hard
photons.}  Therefore, these nonperturbative effects which are not power
suppressed constitute less than five percent uncertainty in the theoretical
prediction for the $\bar B\to X_s\,\gamma$ decay rate.

There is no OPE that allows one to parametrize nonperturbative effects from the
photon coupling to light quarks in terms of $B$ meson matrix elements of local
operators.  Given this, it is perhaps not surprising that nonperturbative
effects that come from the photon coupling to the charm quark contain $B$ meson
matrix elements of local operators that are suppressed by $(\Lambda_{\rm
QCD}/m_c)^2$ rather than $(\Lambda_{\rm QCD}/m_b)^2$.  Recently, Voloshin
identified such a nonperturbative correction to the $\bar B\to X_s\,\gamma$
decay rate \cite{Volo}.  This contribution arises from the interference of
$O_2$ with $O_7$ corresponding to the diagram shown in Fig.~2, and can be
studied using the operator product expansion.  

\begin{figure}[t]  
\centerline{\epsfysize=3truecm \epsfbox{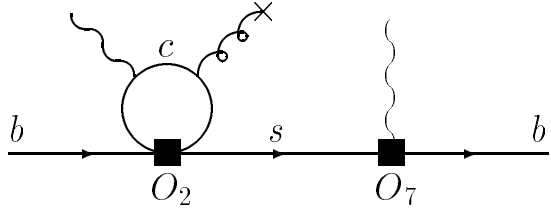}}
\caption[2]{Feynman diagram that gives rise to $T_{27}$ in Eq.~(\ref{mxel}).  
Interchange of the photon and gluon couplings to the charm loop is understood.}
\end{figure}

For a sufficiently heavy charm quark, nonperturbative corrections to the
contribution of the interference of $O_2$ and $O_7$ to the decay rate can be
computed from the discontinuity of the diagram in Fig.~2.  Analogous diagrams
with more gluons give effects suppressed by additional powers of $\Lambda_{\rm
QCD}/m_c$.  Denoting the gluon momentum by $k$, we work to all orders in
$k\cdot q/m_c^2$ since the photon momentum $q$ is of order $m_b$, but neglect
terms of order $k\cdot q/m_b^2$, $k^2/m_{c,b}^2$, and $m_s/m_b$.  The result of
the loop integration is 
\begin{equation}\label{mxel}
T_{27} = - {1\over2m_B}\, \langle\bar B(v)|\, \bar b\;
  m_b\,\sigma^{\nu\rho}q_\rho\,
  {m_b v\!\!\!\slash-q\!\!\!\slash\over(m_bv-q)^2+i\epsilon}\, 
  \gamma^\mu(1-\gamma_5)\, I_{\mu\nu}\, b\, |\bar B(v)\rangle \,.
\end{equation}
$I_{\mu\nu}$ is a complicated operator involving all powers of 
$(q\cdot iD)/m_c^2$.  It is given by
\begin{equation}\label{Imunu}
I_{\mu\nu} = \bigg({e\over16\pi^2}\bigg)^2\, {2\over9m_c^2}\,
  \Bigg[ \sum_{n=0}^\infty {3\cdot2^{n+3}\,[(n+1)!]^2\over(2n+4)!}\,
  \bigg({-q\cdot iD\over m_c^2}\bigg)^{\!n} \Bigg]\,
  \varepsilon_{\mu\nu\lambda\beta}\, q^\beta q_\eta\, g_sG^{\lambda\eta} \,.
\end{equation}
Here $G^{\lambda\eta}$ is the gluon field strength tensor and $D$ denotes
the covariant derivative.  The contribution of $T_{27}$ to the
$\bar B\to X_s\,\gamma$ decay rate is given by Eq.~(\ref{77spec}) with
$C_7^2\,T_{77}$ replaced by $2C_2C_7\,T_{27}$.  

For the leading $n=0$ term in Eq.~(\ref{Imunu}), the matrix element in 
Eq.~(\ref{mxel}) can be computed using the identity \cite{FaNe}
\begin{equation}
{1\over2m_B}\, \langle\bar B(v)|\, \bar b\, \Gamma\, g_sG_{\alpha\beta}\, b\,
  |\bar B(v)\rangle = {\lambda_2\over8}\, {\rm Tr}\, 
  \{\Gamma\,(1+v\!\!\!\slash)\, \sigma_{\alpha\beta}\, (1+v\!\!\!\slash)\} \,,
\end{equation}
valid for any Dirac structure $\Gamma$.  The ratio of the decay rate 
from the $n=0$ term in Eq.~(\ref{Imunu}) to that from $O_7$ is 
\begin{equation}\label{res}
{\delta\Gamma(\bar B\to X_s\,\gamma) \over \Gamma(\bar B\to X_s\,\gamma)} = 
  - {C_2\over9C_7}\, {\lambda_2\over m_c^2} \,.
\end{equation}
The measured $B^*-B$ mass splitting gives $\lambda_2=0.12\,{\rm GeV}^2$.  
Using this value for $\lambda_2$ and the values of $C_2=1.11$ and $C_7=-0.32$
in Ref.~\cite{BMMP}, Eq.~(\ref{res}) implies that this $O_2-O_7$ interference
is about a three percent effect.  This is an order of magnitude larger than the
perturbative estimate of the contribution from the interference of $O_2$ and
$O_7$ to the $\bar B\to X_s\,\gamma$ decay rate (which contains a gluon in the
final state).

The contribution of all terms in $I_{\mu\nu}$ to 
${\rm Im}\,T_{27}$ is 
\begin{eqnarray}\label{sum1}
{\rm Im}\,T_{27} &=& {1\over2m_B}\, \sum_{n=0}^\infty 
  {a_n\, (-1)^n\, m_b^{n+3} \over m_c^{2n+2}}\,
  \widehat q^{\mu_1}\ldots \widehat q^{\mu_n} \\* 
&&\times \langle\bar B(v)|\,\bar b\, \Gamma^{\alpha\beta}(\widehat q,v)\, 
  (iD_{\{\mu_1}\ldots iD_{\mu_n\}}\, g_sG_{\alpha\beta})\, 
  b\,|\bar B(v)\rangle\, \delta(\widehat q\cdot v-1/2) \,, \nonumber
\end{eqnarray}
where $\widehat q=q/m_b$. $\Gamma^{\alpha\beta}$ and $a_n$ are dimensionless
and can be deduced from Eqs.~(\ref{mxel}) and ~(\ref{Imunu}).  The indices
$\mu_1\ldots\mu_n$ are symmetrized, since they are dotted into $\widehat
q^{\mu_1}\ldots\widehat q^{\mu_n}$.  Note that the derivatives
$iD_{\{\mu_1}\ldots iD_{\mu_n\}}$ act on the gluon field $G_{\alpha\beta}$, and
are determined by the spacetime dependence of the chromomagnetic field in the
$B$ meson.  The $n=0$ term in Eq.~(\ref{sum1}) is a special case in that the
$\langle\bar B(v)|\,\bar b\,\Gamma^{\alpha\beta}\,g_sG_{\alpha\beta}\,b\,|\bar
B(v)\rangle$ matrix element is known from the $B^*-B$ mass splitting.  The
$n=1$ matrix element vanishes by the equations of motion \cite{ucla}.  The
$n>1$ terms in Eq.~(\ref{sum1}) depend on an infinite series of unknown matrix
elements.  Estimating $\langle\bar B(v)|\,\bar b\,\Gamma^{\alpha\beta}(\widehat
q,v)\, (iD_{\mu_1}\ldots iD_{\mu_n}\,g_sG_{\alpha\beta})\,b\,|\bar
B(v)\rangle/(2m_B) \sim(\Lambda_{\rm QCD})^{n+2}$, we see that the $n>1$ terms
are ``suppressed" compared to the $n=0$ term considered by Voloshin only by
powers of $m_b\,\Lambda_{\rm QCD}/m_c^2$.  

In the limit where $m_c$ is fixed and $m_b\to\infty$, the higher order terms in
Eq.~(\ref{sum1}) become successively more important and the expansion we have
made is clearly inappropriate.  (The whole sum in Eq.~(\ref{sum1}) is, up to
logarithms, of order $\Lambda_{\rm QCD}/m_b$.)  In the limit where $m_b/m_c$ is
held fixed and both masses become very large, the $n\geq1$ terms in
Eq.~(\ref{sum1}) are suppressed by powers of $\Lambda_{\rm QCD}/m_c$.  Then the
$n=0$ result, which is of order $\Lambda_{\rm QCD}^2/m_c^2$, dominates the sum.
In the physical world, $m_b\,\Lambda_{\rm QCD}/m_c^2\sim0.6$ (an equally
reasonable estimate would be $E_\gamma^{\rm max}\,\bar\Lambda/m_c^2$, which is
also about 0.6).  From Eq.~(\ref{Imunu}) we see that $a_1/a_0=4/15$,
$a_2/a_0=3/35$, $a_3/a_0=16/525$, etc., and asymptotically
$a_n/a_0\to3\sqrt\pi/(2^{n+1}\,n^{3/2})$ as $n\to\infty$.  The values of
$a_n/a_0$, $n=1,2,\ldots$, are small.  This together with the asymptotic
formula for large $n$ suggests that the $n\geq1$ terms in Eq.~(\ref{sum1}) do
not introduce a nonperturbative uncertainty greater than the value of the
leading $n=0$ term.  Nonperturbative effects from the interference of $O_1$
with $O_7$ are expected to be smaller.

Near the photon end-point region, another set of corrections become large. 
Expanding factors of $iD$ that occur in the denominator of the strange quark
propagator (these were neglected in Eq.~(\ref{mxel})) yields corrections
suppressed by powers of $m_b$.  However, these corrections are proportional to
derivatives of the delta function $\delta(E_\gamma-m_b/2)$, and they become as
important as those in Eq.~(\ref{sum1}) in the end-point region.  

Consider next the contribution to the $\bar B\to X_s\,\gamma$ decay rate coming
from the square of $C_1O_1+C_2O_2$.  Diagrams like that in Fig.~2 should give a
smaller nonperturbative contribution to the decay rate than the interference of
$O_2$ with $O_7$ (i.e., these are order $\Lambda_{\rm QCD}^4/m_c^4$ instead of
order $\Lambda_{\rm QCD}^2/m_c^2$).  But we know in this case that there is a
contribution to the $\bar B\to X_s\,\gamma$ decay rate from $\bar B\to
X_s\,J/\psi$ followed by $J/\psi\to\gamma\,X$, which is much larger than the
perturbative calculation of the effect of $(C_1O_1+C_2O_2)^2$.  The combined
branching ratio for this process is about $10^{-4}$, while the perturbative
estimate of the contribution of $(C_1O_1+C_2O_2)^2$ is less than $10^{-5}$. 
This might not present a serious difficulty for the comparison of experiment
with theory, since the process $\bar B\to X_s\,J/\psi$ followed by
$J/\psi\to\gamma\,X$ does not favor hard photons, and in any case it can be
treated as a background and subtracted away.  Further work on this issue is
warranted.

In this letter we examined uncertainties in the theoretical prediction for the
weak radiative decay rate of $B$ mesons into hard photons that come from
nonperturbative strong interaction physics.  We focused on effects that arise
from photon couplings to light quarks and to charm quarks.  For hard photons
the first of these sources of theoretical uncertainty is less than five
percent.  This is smaller than the uncertainty in the Wilson coefficient
$C_7(m_b)$ from uncalculated order $\alpha_s^2$ terms in its perturbative
expansion.  For the photon coupling to the charm quark, more work is needed to
decide the size of the theoretical uncertainty associated with nonperturbative
effects.

The present experimental data on $\bar B\to X_s\,\gamma$ focuses on photon
energies in the region $E_\gamma\gtrsim2.2\,$GeV \cite{CLEO}.  For comparison
with this data, the largest theoretical uncertainty is from the contribution of
higher dimension operators to the time ordered product $T_{77}$ which become
more important in the end-point region.  This uncertainty would be
substantially smaller if the photon energy cut were reduced.

\acknowledgements
Similar work has been done by Grant, Morgan, Nussinov, and Peccei \cite{ucla}. 
We thank them for discussing their results with us prior to publication.  We
thank Gerhard Buchalla, Gino Isidori, Anton Kapustin, David Politzer, Soo-Jong
Rey, Ira Rothstein, and Hitoshi Yamamoto for useful conversations.  
Z.L. and M.B.W. were supported in part by the U.S.\ Dept.\ of Energy under 
grant no.\ DE-FG03-92-ER~40701.  
L.R. was supported in part by the DOE under cooperative agreement number
DE-FC02-94ER40818, an NSF Young Investigator award, an Alfred P. Sloan
Foundation Fellowship, and a DOE Outstanding Junior Investigator award.

{\tighten

} 


\begin{references}


\bibitem{CLEO}
CLEO Collaboration, M.S. Alam {\it et al.}, Phys. Rev. Lett. 74 (1995) 2885.

\bibitem{old}
S. Bertolini {\it et al.}, Phys. Rev. Lett. 59 (1987) 180; \\
N.G. Deshpande {\it et al.}, Phys. Rev. Lett. 59 (1987) 183.

\bibitem{np}
See, e.g., B. Grinstein and M.B. Wise, Phys. Lett. B201 (1988) 274; \\
W-S. Hou and R.S. Willey, Phys. Lett. B202 (1988) 591; \\
S. Bertolini {\it et al.}, Nucl. Phys. B353 (1991) 591;  \\
J.L. Hewett, SLAC-PUB-6521 [hep-ph/9406302], and references therein.

\bibitem{GSW}
B. Grinstein {\it et al.}, Phys. Lett. B202 (1988) 138;
  Nucl. Phys. B339 (1990) 269; \\
M. Misiak, Phys. Lett. B269 (1991) 161; Nucl. Phys. B393 (1993) 23; \\
R. Grigjanis {\it et al.},  Phys. Lett. B213 (1988) 355 [(E) B286 (1992) 413].

\bibitem{LLO}
M. Ciuchini {\it et al.}, Phys. Lett. B316 (1993) 127; \\
M. Ciuchini {\it et al.}, Nucl. Phys. B421 (1994) 441.

\bibitem{BMMP}
A.J. Buras {\it et al.}, Nucl. Phys. B424 (1994) 374.

\bibitem{AG}
A. Ali and C. Greub, Z. Phys. C49 (1991) 431; Phys. Lett. B259 (1991) 182.

\bibitem{KLP}
A. Kapustin {\it et al.},  Phys. Lett. B357 (1995) 653.

\bibitem{lql}
N. Pott, Phys. Rev. D54 (1996) 938; \\
C. Greub {\it et al.}, Phys. Rev. D54 (1996) 3350.

\bibitem{nlomatch}
K. Adel and Y.P. Yao, Phys. Rev. D49 (1994) 4945.

\bibitem{nlo}
K. Chetyrkin {\it et al.}, ZU-TH-24-96 [hep-ph/9612313]; \\
M. Misiak and M. M\"unz, Phys. Lett. B344 (1995) 308.

\bibitem{FLS}
A.F. Falk {\it et al.}, Phys. Rev. D49 (1994) 3367; \\
I. Bigi {\it et al.}, in {\it The Fermilab Meeting}, Proc.\ of the Annual
Meeting of the DPS of the APS, ed.\ C. Albright {\it et al.} 
(World Scientific, Singapore, 1993), p. 610.

\bibitem{Matthias}
M. Neubert, Phys. Rev. D49 (1994) 4623; \\
I.I. Bigi {\it et al.}, Int. J. Mod. Phys. A9 (1994) 2467.

\bibitem{AKZL}
A. Kapustin and Z. Ligeti, Phys. Lett. B355 (1995) 318; \\
R.D. Dikeman {\it et al.}, Int. J. Mod. Phys. A11 (1996) 571. 

\bibitem{Volo}
M.B. Voloshin, TPI-MINN-96/30-T [hep-ph/9612483].

\bibitem{CGG}
J. Chay {\it et al.}, Phys. Lett. B247 (1990) 399; \\
M. Voloshin and M. Shifman, Sov. J. Nucl. Phys. 41 (1985) 120.

\bibitem{ed}
E. Witten, Nucl. Phys. B120 (1977) 189;  \\
C.H. Llewellyn Smith, Phys. Lett. B79 (1978) 83;  \\
K. Koller {\it et al.}, Z. Phys. C2 (1979) 197.

\bibitem{lore}
J.F. Owens, Rev. Mod. Phys. 59 (1987) 465; and references therein.

\bibitem{FaNe}
A.F. Falk and M. Neubert, Phys. Rev. D47 (1993) 2965.

\bibitem{ucla}
A.K. Grant {\it et al.}, UCLA/97/TEP/5 [hep-ph/9702380].


\end{references}
\end{document}